\documentclass[a4paper,11pt]{article}
\pdfoutput=1 

\usepackage{jheppub} 
            
\usepackage{bm,amssymb,slashed,graphicx,multirow,soul,mathtools,xspace,array,comment}  
\usepackage{float}                   
\allowdisplaybreaks
\usepackage{ bbold }
\usepackage{color}
\usepackage{subfigure}
\usepackage[printonlyused]{acronym}
\usepackage{hyperref} 
\usepackage{overpic}
\acrodef{PDG}[PDG]{Particle Data Group}
\acrodef{OPE}[OPE]{Operator Product Expansion}
\acrodef{FCNC}[FCNC]{flavour-changing neutral current}
\acrodef{RHC}[RHC]{right-handed currents}
\acrodef{SM}[SM]{Standard Model}
\acrodef{NP}[NP]{New Physics}
\acrodef{MFV}[MFV]{Minimal Flavour Violation}
\acrodef{SD}[SD]{short-distance}
\acrodef{LD}[LD]{long-distance}
\acrodef{DA}[DA]{distribution amplitude}

\newcommand{\state}[1]{|#1\rangle}

\newcommand{\matel}[3]{\langle #1|#2|#3\rangle}
\newcommand{\al}{\alpha}
\newcommand{\be}{\beta}
\newcommand{\ga}{\gamma}
\newcommand{\de}{\delta}
\newcommand{\la}{\lambda}
\newcommand{\eps}{\epsilon}

\newcommand{\GeV}{\,\mbox{GeV}}

\newcommand{\ORD}{{\cal O}}

\newcommand{\Cdot}{\!\cdot \!}
\newcommand{\mi}{\!-\!}

\newcommand{\FIG}{Fig.~}

\newcommand{\SEC}{Sec.~}

\newcommand{\APP}{App.~}

\newcommand{\EQ}{Eq.~}
\newcommand{\EQs}{Eqs.~}

\newcommand{\DEL}{\delta}

\newcommand{\phiB}{{\Phi_{B}}}
\newcommand{\phiBda}{{\Phi^\dagger_{B}}}
\newcommand{\phiBpl}{{\Phi_{B^-}}}
\newcommand{\momB}{{p}_\phiB}
\newcommand{\mphiB}{m_{\phiB}}

\newcommand{\rr}{r}

\newcommand{\lepB}{B^- \to \ell^- \bar \nu}
\newcommand{\lepBz}{B \to \ell \bar \nu}

\newcommand{\Hpz}{{\cal J}^{(0)} _B}
\newcommand{\Hpzda}{{\cal J}^{(0)\,\dagger}_B}
\newcommand{\ZB}{Z_B}
\newcommand{\ZBp}{{\cal Z}_B^{(0)}}
\newcommand{\pot}{\varphi}
\newcommand{\JBU}{{{\cal J}}^{(D)}_B}
\newcommand{\JBV}{ {\cal J}_B}
\newcommand{\KB}{K}
\newcommand{\JDirac}{{\cal I}}
\newcommand{\JDiracO}[1]{{\JDirac_{#1} }}
\newcommand{\JDiracB}[2]{{\JDirac^{#1}_{#2} }}

\definecolor{violet}{rgb}{0.94, 0.2, 0.8}
\definecolor{lightblue}{rgb}{0.39, 0.58, 1.00} 
\definecolor{lightgreen}{rgb}{0.1, 0.73, 0.33}

\numberwithin{equation}{section}

\title{ \boldmath Resolving Charged Hadrons in QED - \\
 Gauge Invariant Interpolating Operators}
\author[1]{Saad Nabeebaccus,}
\author[2]{Roman Zwicky}

\affiliation[1]{Universit\'e Paris-Saclay, CNRS/IN2P3, IJCLab, 91405 Orsay, France}
 \affiliation[2]{Higgs Centre for Theoretical Physics, School of Physics and Astronomy, University of Edinburgh, Edinburgh EH9 3JZ, Scotland}

\emailAdd{saad.nabeebaccus@ijclab.in2p3.fr}
\emailAdd{roman.zwicky@ed.ac.uk}

\abstract{
Standard interpolating operators for charged mesons, e.g. $J_{B} =
\bar b i \gamma_5 u$ for $B^-$, are 
not gauge invariant in QED and therefore problematic for perturbative methods.  
We propose a gauge invariant interpolating operator by adding an auxiliary charged scalar $\Phi_B$, 
${\cal J}_{B}^{(0)} = J_B \, \Phi_B$,  which reproduces all the universal soft and collinear logs.
The modified LSZ-factor is shown to be infrared finite which is a necessary condition for validating the approach.  At  ${\cal O}(\alpha)$, this is equivalent to a specific Dirac dressing of charged operators.
A generalisation thereof, using iterated integrals, 
establishes the equivalence to all orders and provides a transparent alternative viewpoint. 
The method is discussed by the example of the leptonic decay $B^- \to \ell^- \bar \nu$ for which a numerical study is to follow.  The  formalism itself is valid for any spin, flavour and set of final states  (e.g.  $B^- \to \pi^0  \ell^- \bar \nu$).
}

\begin{document}

\toccontinuoustrue

\maketitle

\flushbottom

\setcounter{tocdepth}{3}
\setcounter{page}{1}
\pagestyle{plain}


\section{Introduction}
\label{sec:intro}

Scalar QED or the point-like approximation is a consistent framework which works numerically well in many cases, but precision in CKM matrix elements and testing of lepton flavour universality \cite{Bifani:2018zmi,LHCb:2021lvy} are calling 
for stucture-dependent computations.\footnote{Scalar QED is  sufficient in reducing 
the $R_K$ uncertainty to $\ORD(1\%)$  
as there are no enhanced $\ORD(\al) \ln \frac{m_\ell}{m_B}$ logarithms (logs)  beyond scalar QED 
\cite{Isidori:2020acz} and charmonium resonances are under control for $q^2 < 6\GeV^2$ 
\cite{Isidori:2022bzw}.}  
  Approaches  include chiral perturbation theory (ChPT) 
 \cite{Cetal01,CGH08,Descotes-Genon:2005wrq}, 
soft collinear effective theory (SCET)  \cite{BBS17,Beneke:2019slt}   and lattice 
Monte Carlo simulations (with a range of methods  to contain the massless photon in finite volume \cite{Carrasco:2015xwa,Endres:2015gda,Lucini:2015hfa,Feng:2018qpx} and 
applications thereof in     
\cite{Lubicz:2016xro,Giusti:2017dwk,Hansen:2018zre,DiCarlo:2019thl,Clark:2022wjy}). ChPT applies to low energy physics and  might be viewed as a successful expansion around the point-like approximation 
where the challenge relies in determining the finite counterterms.  
In SCET, mesons are described by light-cone distribution amplitudes which necessitate 
the introduction of process dependent Wilson lines \cite{Beneke:2021pkl,Beneke:2022msp}. In lattice Monte Carlo simulations, hadrons
are described by either gauge variant or  gauge invariant interpolating operators (on which 
we comment at the end of \SEC\ref{sec:inter}).
No method exists for perturbative approaches  
with interpolating operators 
for charged mesons.\footnote{The approach we have in mind is QCD sum rules (cf. 
\APP \ref{app:QCDSR}) where the 
Lehmann Symanzik Zimmerman (LSZ) approach (e.g. \cite{Duncan:2012aja}) is implemented 
via semi-global quark-hadron duality \cite{Shifman:1978bx,Shifman:2000jv}} 
It is the aim of this paper to fill this gap.  

This work is organised as follows. 
In \SEC\ref{sec:prelim}, some preliminary materials 
such as the  problem of gauge variance 
and  the universal infrared (IR) logs are discussed.
In \SEC\ref{sec:phiB}, we introduce the idea of the long distance
$B$-meson as part of the modified gauge invariant interpolating operator.  
In \SEC\ref{sec:LSZ}, it is  shown and argued 
that  its LSZ-factor is IR-finite. 
In \SEC\ref{sec:Dirac}, we establish the connection with the Dirac dressing at $\ORD(\al)$, 
and a generalisation using iterated integrals establishes the connection to all orders. 
 The paper ends with conclusions and discussions in \SEC\ref{sec:conclusions}.
In \APP\ref{app:necessity}, we comment on the necessity of gauge invariant interpolating operators, 
in connection with  the  hard photon approach to $\lepB \ga$. In \APP\ref{app:QCDSR}, we briefly  
review the  QCD sum rules approximation to the LSZ approach.

\section{Preliminaries}
\label{sec:prelim}

\subsection{Gauge variance of the QCD interpolating operator}
\label{sec:gauge}

Let us first discuss the problem in a pedestrian way through the example of 
a leptonic decay of the type  $\lepB$. In QCD, mesons such as the $B^-$  
are interpolated  by \footnote{In QCD, it is advantageous to include  $(m_b+m_u)$ as a 
prefactor in $J_B$ as it becomes a renormalisation group invariant quantity then. However, in QED,
this will be of no use since this property is spoiled by the fact that the $m_q$ renormalises as $e^2 Q_q^2$ and thus we omit the masses in $J_B$.}
\begin{equation}
\label{eq:JB}
J_{B}  = 
\bar b i \ga_5 u  \;, \quad  \ZB = \matel{B^-}{J_B(0)}{0}  = \frac{m_B^2 f_B}{ (m_b+m_u)} \;, 
\end{equation}
where $\ZB$  plays  the r\^ole of the LSZ-factor, This means  that   the matrix element 
\begin{equation}
\label{eq:LSZ}
 \matel{0}{ O(x)}{B^-} =  \frac{1 }{\ZB} \lim_{p^2 \to m_B^2}  ( m_B^2-p^2) \Pi_{OB}(p^2)\;,
 \end{equation}
 can be extracted from the correlation function
\begin{equation}
\label{eq:OB}
 \Pi_{OB}(p^2) = i \int_x  e^{i xp}  \matel{0}{T O(x) J_B(0)  }{0} =
\int_0^\infty   \frac{ds}{2 \pi i}  \frac{\text{disc}_s \Pi_{OB}(s)}{s- p^2 -i0} = 
\frac{\matel{0}{ O(x)}{B^-}  \ZB}{m_B^2-p^2} + \dots \;,
\end{equation}
since the latter satisfies a dispersion relation as indicated.  In \eqref{eq:OB}, $\text{disc}_s \Pi(s) =  \Pi(s+i0) -  \Pi(s-i0)$ is the discontinuity across the real line, 
the dots stand for higher states in the spectrum and  $\ZB^* = \ZB$ has been assumed.
The shorthands $\int_x = \int  d^4 x$ and  $xp = x \cdot p$, used above, are occasionally  assumed hereafter.

When one considers QED, the picture is  fundamentally disturbed in perturbation theory since the operator 
$J_{B} $ (or $\ZB$)
ceases to be  gauge invariant\,\footnote{c.f. \APP\ref{app:exact} for comments on the non-perturbative case.}

\begin{equation}
\label{eq:GV}
J_{B}  \to e^{- i \la Q_{B}} J_{B}  \;, 
\end{equation}
under 
\begin{equation}
\label{eq:GT}
A \to A - \partial \la \;, \quad  q \to  q e^{i Q_q \la} \;.
\end{equation}
Hence, if there is charge ($Q_B \equiv  Q_{B^-} = Q_b - Q_u = -1$ here and below), as is well-known, $f_B$  ceases to be an observable once QED corrections are considered (e.g. \cite{Gasser:2010wz} for a discussion for $f_\pi$). 
Note that \eqref{eq:GT}
implies the $D =  \partial +  i A$ convention with the electric charge absorbed into 
the photon field $A$  such that  $n$  photon fields correspond to $\ORD(e^n)$. 
 
 \subsection{Universal  infrared sensitive logs in QED}
\label{sec:universal}

One of the main features  of QED is the appearance of IR sensitive logs which 
can overcome the small coupling constant $\al = e^2/4 \pi \approx 1/137$
and a fair amount is known about them. 
In particular, their physics is dictated and controlled by gauge invariance and thus complications 
are to be expected 
 when  gauge invariance is not manifest. 
One may distinguish two classes of  logs:  first, the
so-called (hard) collinear logs, which arise from (charged) 
particles $\ell^-$ with small mass, allowing for collinearity with the photon up to $\ORD(\hat{m}_\ell^2)$
and resulting in sizeable   $\ORD(\al) \ln \hat{m}_\ell$-terms (hatted quantities are divided by $m_B$ hereafter).
Second, there are  soft and soft-collinear logs of the form 
$\ORD(\al) \ln \hat{m}_\ga$ and $\ORD(\al) \ln \hat{m}_\ga \ln \hat{m}_\ell$ which are true IR divergences 
but turn into $\ORD(\al) \ln \hat{\DEL} $ and $\ORD(\al) \ln \hat{\DEL} \ln \hat{m}_\ell$ when the  photon emission, 
of  $E_\ga < \DEL$, is added.  
Aspects of resummation of these logs are known \cite{Yennie:1961ad,Kuraev:1985hb} 
 but it is rather  their 
degree of independence of the structure (or \emph{universality}) 
that is of interest to us.  
Soft and soft-collinear logs are universal since the soft photons cannot resolve the structure of the
mesons. For hard-collinear logs, the situation is not as transparent  as  photon energies can be as large as the kinematics 
allow for.  It turns out that  gauge invariance and the KLN-theorem 
(cancellation of all real and virtual  IR-logs in the $m_\ell \to 0 $ limit, in the photon inclusive case)  
are sufficient premises to show that there are no further hard-collinear logs of the form  
$\ORD(\al)  \ln \hat{m}_\ell$ \cite{Isidori:2020acz}.  In other words, structure-dependent $\ln \hat{m}_\ell$ terms are subleading, either in $\ORD(\al)  \hat{m}_\ell^2 \ln \hat{m}_\ell$ or $\ORD(\al^2)  \ln \hat{m}_\ell$, both of which are
negligible.\footnote{This does not forbid other relevant and interesting structure-dependent effects such 
 as the $m_b/\Lambda_{\textrm{QCD}}$-enhancement  found in 
$B_s \to \mu^+\mu^-$ [8]. 
In $B^- \to \ell^- \bar \nu$ such effects might be absent because the equation of motion 
of the lepton, that is the helicity suppression,  work out in different ways.}

However,  when the leading order (LO) process is $\ORD(m_\ell^2)$-suppressed, 
the theorem does not apply as then real and virtual contributions  are not linked by the KLN-theorem \cite{Zwicky:2021olr}. This is precisely the case for  $\lepB$ enabled by V-A interactions, 
as in the  Standard Model (SM). 
These non-universal structure-dependent  logs  complicate the investigation of the validity of the approach.
However, we can easily bypass this issue by resorting to an  S-P interaction 
\begin{equation}
\label{eq:LW}
{\cal L}_{ \lepB}^{\small (S-P)} = g_{S-P} \bar u (1\mi \ga_5   )b \, \bar \ell (1\mi \ga_5   ) \nu \;,
\end{equation}
which is not $\ORD(m_\ell^2)$-suppressed. Hence, the hard-collinear logs are universal and, 
in particular, reproducible from the splitting function since the KLN-theorem applies \cite{Zwicky:2021olr}. 
We stress that the reasoning for choosing an S-P interaction that it allows us to test whether or not 
our method is capable of  reproducing universal collinear logs.
In summary, quoting almost verbatim from \SEC 3.2.1 of that reference, one has
\begin{equation}
\label{eq:leptonic-rate}
 \Gamma( \lepBz (\gamma )) =    \Gamma( \lepBz)^{(0)}(1 + \frac{\al}{4 \pi}\left(F_{\textrm{soft}}(\hat{m}_\ell , 2 \hat{\DEL})   + F_{\textrm{coll}}(\hat{\DEL}) \ln \hat{m}_\ell  + \textrm{non-log}   \right)) \;,
 \end{equation}
where  the soft factor is 
 \begin{equation}
 \label{eq:soft-factor}
F_{\textrm{soft}}(x,y) =  - (4 \frac{1+x^2}{1-x^2}  \ln x^2 + 8)  \ln y \;,
 \end{equation}
 and the (hard) collinear part, reproducible from the splitting function,  reads   
\begin{equation}
\label{eq:coll-factor}
F_{\textrm{coll}} (\hat{\DEL}) = -4 ( \frac{3}{2} - 2 \hat{\DEL}(2 - \hat{\DEL} )) \;.
\end{equation}
This  has been backed up by an explicit computation \cite{NRZ22}.  
We note that in the photon inclusive limit,
$2 \hat{\DEL} \to 1 - \hat{m}_\ell^2$, which implies that $F_{\textrm{coll}} (\hat{\DEL}) \to \ORD(m_\ell^2)$ as required by 
the KLN-theorem since these logs are not suppressed by a factor of $m_\ell^2$ in the rate \eqref{eq:leptonic-rate}.

The essential starting point of this paper is the observation that for 
 the gauge variant interpolating current $J_B$ \eqref{eq:JB}, these universal logs are not reproduced 
 in a perturbative computation.  We shall see how to remedy it and how to interpret it in due course. 

\section{The Long Distance $B$-meson as a Scalar Field $\phiB$}
\label{sec:phiB}

It is well-known that off-shell correlation functions are not gauge invariant, be it in QED or QCD, 
and this is at the heart of the issue raised in the introduction; namely, that the universal IR-logs are 
not reproduced when computing with the interpolating operator $J_B$ \eqref{eq:JB}. 
On the other hand, in scalar QED (point-like approximation), no such problems occur since the external particles can all be put on-shell. It is thus tempting to cure both, gauge invariance and the 
universality problem, by 
introducing a long distance (on-shell) $B$-meson in terms of  a scalar field $\phiB$ of mass $m_B$ as follows
\begin{equation}
\label{eq:JBp} 
 \Hpz  \equiv J_B  \phiB \;, \quad 
\ZBp  \equiv \matel{B^- }{\Hpz}{\phiBpl }  \;.
\end{equation}
The matrix element $\ZBp$ takes on  the r\^ole of the LSZ-factor and  its IR-finiteness is discussed in \SEC\ref{sec:LSZ}.  
The explicit gauge invariance of $\Hpz$ (or $\ZBp)$, as opposed to $J_B$ \eqref{eq:GV},  is  guaranteed 
\begin{equation}
\Hpz \to  e^{i \la ( Q_{\phiB} - Q_{B})}  \Hpz|_{Q_{B} = \,Q_{\phiB}} =  \Hpz \;,
\end{equation}
by choosing  $\state{\phiBpl}$ to have the same charge as $\state{B^-}$;  rendering $\Hpz$ charge neutral.  Our master formula for computing the decay rate is then 
 \begin{equation}
 \label{eq:master}
 \Gamma_{\de}(\lepB (\ga)) = 
 \frac{1}{   |\ZBp|^2  } \times
  \int_{\de} d \Phi_\ga  | \ZBp \, {\cal A}(\lepB (\ga)) |^2   \;, 
 \end{equation}
where both terms, the LSZ-factor $|\ZBp|^2$ and the integrand,   are computed separately, 
and $\int_{\de} d \Phi_\ga $ is the integral over the photon phase space with $E_\ga < \de $ 
(cf. \cite{Isidori:2020acz} for more detail). 
The  amplitude  squared  is given by
\begin{equation}
| {\cal A}(\lepB (\ga)) |^2 = | {\cal A}(\lepB ) |^2 \de(\Phi_\ga)   +|  {\cal A}(\lepB \ga) |^2 \;,
\end{equation}
where $ \de(\Phi_\ga) $ is a delta function in the photon variables as appropriate for the virtual contribution. 
This is the famous Bloch-Nordsieck mechanism at work which
bypasses the QED IR-problem of  charged particles.
As previously mentioned,  the $\phiBpl$-particle can be considered as the long distance version of the $B$-meson 
which splits into its partons at the $\Hpz$-vertex. While being appealing, this idea 
should be met with scepticism at first. Its validation proceeds in several steps. 
The reproduction of the IR-sensitive logs from the diagrams in \SEC\ref{sec:main} (numerator in \eqref{eq:master}), 
the IR-finiteness of the LSZ-factor $\ZBp$ in \SEC\ref{sec:LSZ} (denominator in \eqref{eq:master})
and the reinterpretation in terms non-local operators in  \SEC\ref{sec:Dirac}.

\subsection{The main process}
\label{sec:main}

In this section, we describe how the main process, by  which we mean everything  in \eqref{eq:master}  but  the 
LSZ-factor, is computed. We depart from the following correlation function 
(with $ {\cal L}_W $, a shorthand for the weak Lagrangian, see \eqref{eq:LW})\footnote{Since we invoke 
the Bloch-Nordsieck mechanism, we may ignore  that the virtual and the real emission part of
the $B$-meson is not a well-isolated state.}
\begin{eqnarray}
\label{eq:mainSR}
\Pi^{(\ga)}(p_B^2, \momB^2) &\;=\;& i  \int _x e^{i x \rr}  \matel{ \ell \bar \nu (\ga) }{ T 
   {\Hpz}   (x) (- {\cal L}_W(0) ) 
 }{  \phiB( \momB)}   \nonumber \\[0.1cm]
&\;=\;&  \int \frac{ds}{2\pi i} \frac{ \text{disc}_s[\Pi^{(\ga)}(s, \momB^2) ]}{s-p_B^2-i0} =   \frac{ \ZBp \,  {\cal A}( 
\lepB (\ga))    }{ m_B^2-p_B^2} + \dots \;,
 \end{eqnarray}
 where $r \equiv \momB - p_B$ is introduced in order to distinguish the $\momB$- and the 
$ p_B$-momenta, even though both are to be set on-shell (to $m_B^2$) in the end.\footnote{The $r$-momentum is auxiliary and momenta are to be chosen such that its effect disappears from the final  
result. This is rather straightforward to implement \cite{NRZ22}.}
For $\momB$, this is straightforward 
\begin{equation}
\label{eq:matrix-element-phiB}
\Pi^{(\ga)}(p_B^2, m_B^2)  =     \lim_{\momB^2 \to m_B^2} (m_B^2 -   \momB^2) \,  i^2 \int_{x,z} e^{i (  x \rr - z \momB ) }  \matel{\ell \bar \nu   (\ga) }{ T   \phiBda (z)  {\Hpz}(x) {\cal L}_{\textrm{W} }(0) }{0}   \;,
\end{equation}
since it plays the r\^ole of an elementary particle. The formal definition of the matrix element (times the LSZ-factor) is then obtained from \eqref{eq:mainSR} as 
\begin{equation}
\label{eq:formal}
\ZBp \,  {\cal A}( \lepB (\ga))  = \lim_{p_B^2 \to m_B^2} ( m_B^2 - p_B^2) \Pi^{(\ga)}(p_B^2, m_B^2) \;.
\end{equation}
We stress  that \eqref{eq:formal} serves only as a formal definition 
of the matrix element since in practice, as is well-known, it is impossible to extract a bound state pole 
with  perturbative methods, since bound states are non-perturbative.   
\begin{centering}
\begin{figure}
\includegraphics[width=1.0\linewidth]{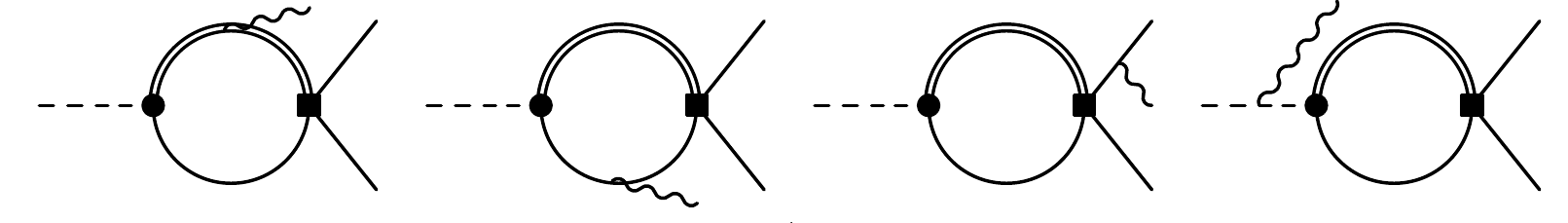}  
	\caption{\small  Diagrams contributing to $\Pi^{(\ga)}(p_B^2,\momB^2)$ in \eqref{eq:mainSR} 
	(i.e. the radiative or real emission part). The last diagram is specific to the $\phiB$-particle.}
	\label{fig:dia-real-nom}
\end{figure}
\end{centering}
\begin{centering}
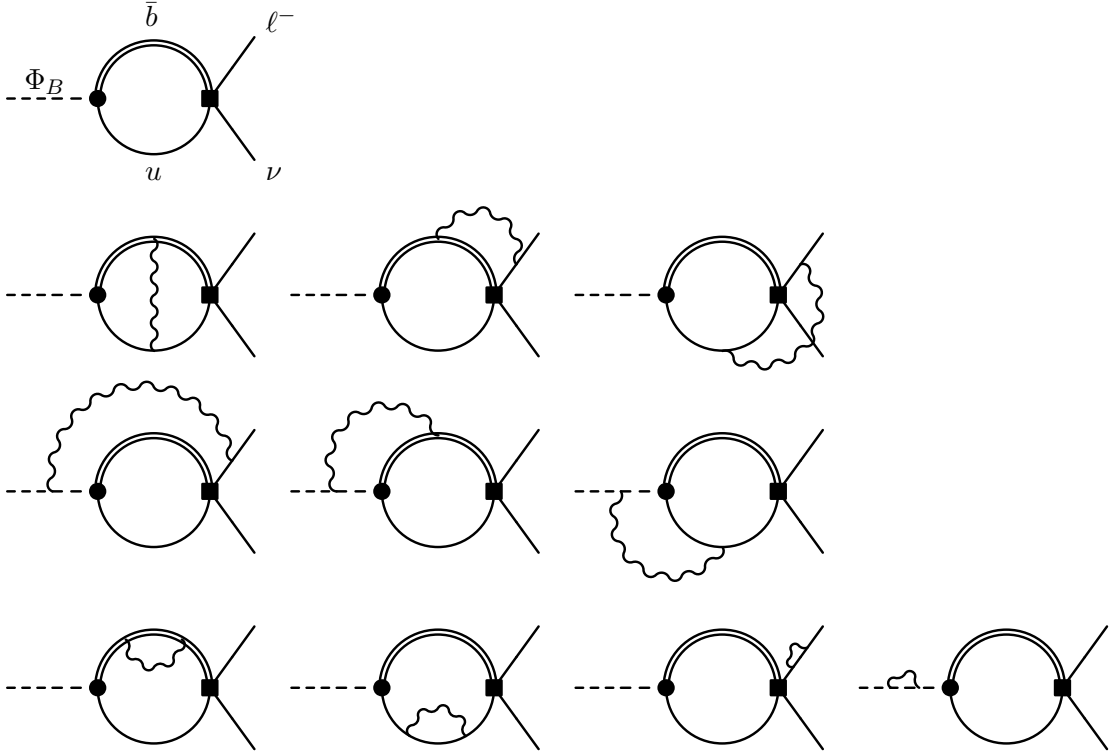
\begin{figure}[h!]
\begin{overpic}[width=1.0\linewidth]{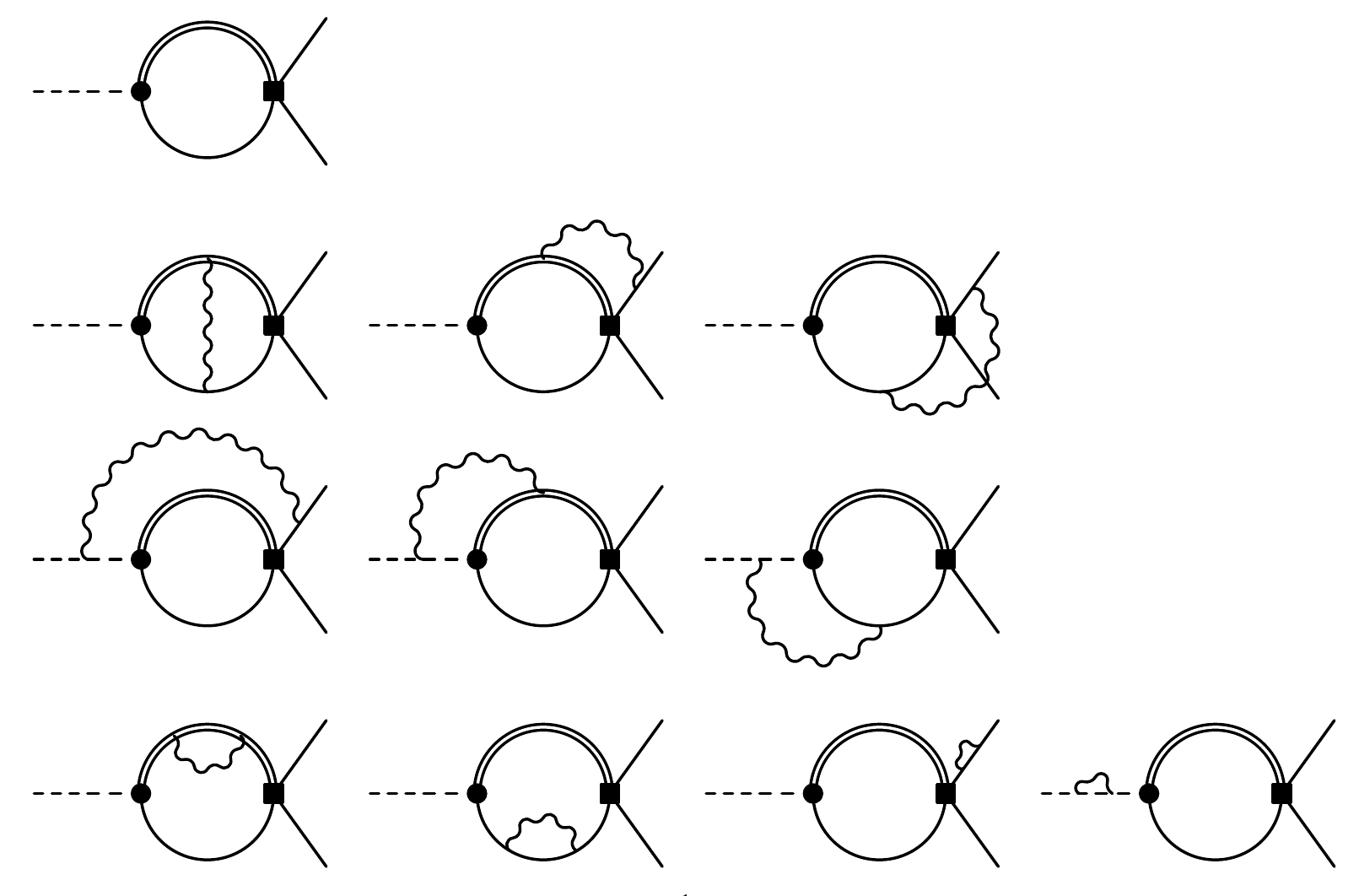}  
\put(4,60){$\phiB$}
\put(14.5,65.5){$\bar b$}
\put(14.5,52){$u$}
\put(25,65){$\ell^-$}
\put(25,52){$\nu$}
\end{overpic}
	\caption{\small  Diagrams contributing to $\Pi(p_B^2,\momB^2)$ in \eqref{eq:mainSR} 
	(ie. the non-radiative part, hence no $(\ga)$ superscript). Top line is the LO diagram and the third line and the last diagram are specific to the $\phiB$-particle.}
	\label{fig:dia-virt-nom}
\end{figure}
\end{centering}

The real emission diagrams are depicted 
in \FIG\ref{fig:dia-real-nom} and the computation of their discontinuities is  straightforward. 
In practice, the main challenge  is to compute the discontinuity of the virtual diagrams of the type shown 
in \FIG\ref{fig:dia-virt-nom}.  
We have performed this task by using Cutkosky rules as the virtual diagrams involve two loops and have 
a considerable number of scales.  The results, with more details to be reported in \cite{NRZ22}, are the following:
\begin{enumerate} 
\item All universal collinear logs \eqref{eq:leptonic-rate} are reproduced \emph{separately} for the
 virtual and  the real rates. They originate from the sum of all  $b \bar u$-cuts as 
 the sum of all  $b \bar u \gamma$-cuts is free from collinear logs (in the S-P case).  
\item The universal   soft (and soft-collinear) logs, given in \eqref{eq:soft-factor}, are equally 
reproduced and emerge as $\ln \hat{\de}$ and $\ln \hat{\de} \ln \hat{m}_\ell$ terms respectively.\footnote{Of course, the real and virtual rates contain soft-divergences, 
which have to be regularised (e.g. dimensional or photon mass regularisation).  The split of real and virtual soft divergences is equally universal in the sense that  
$\ln \de \to \ln m_\ga|_{\text{virtual}} -  \ln m_\ga|_{\text{real}}$ restores $\ln m_\ga$-terms.}   
In order to reproduce  the  soft logs the auxiliary $\Phi_B$ is crucial. Its on-shellness 
gives rise to the correct soft-structure in the integrand.
\item Unphysical IR divergences in $\ln m_u$ and $\ln m_\ga$, of the  collinear and soft type, 
cancel for the sum of all cuts. 
We note that this must be the case since 
the corresponding momentum $p_B$ is off-shell and IR-finiteness follows from the 
Kinoshita-Poggio-Quinn theorem \cite{Muta:1998vi}.
\end{enumerate} 
Let us remark to this end that taking all the cuts is what is usually done in the case of 
virtual QCD computations. However, in the case of hard photon emission, $\lepB \ga$, 
where the photon is energetic (or non-soft), the procedure has been a different one 
in the literature for many decades. Why this   is a valid procedure is explained in 
\APP\ref{app:hard}.

\subsection{The LSZ-factor is infrared finite}
\label{sec:LSZ}

\begin{centering}
\begin{figure}[h!]
\includegraphics[width=1.0\linewidth]{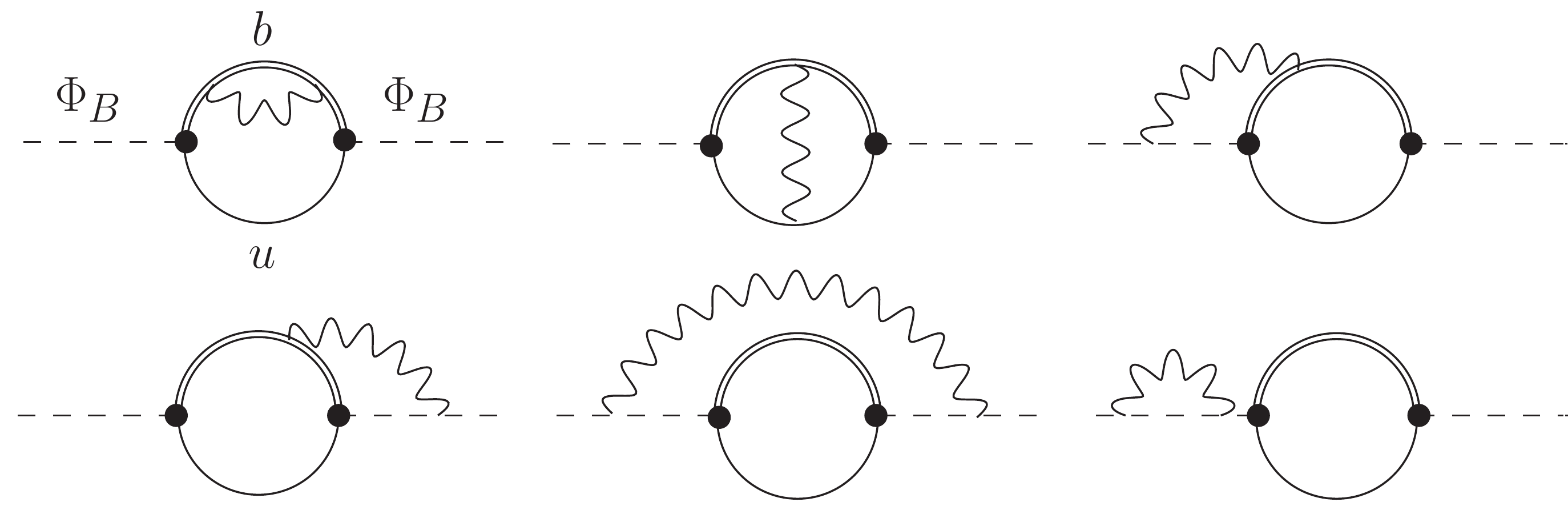}  
	\caption{\small  Diagrams contributing to $C(p_B^2, \momB^2)$ \eqref{eq:C}, 
	that is $|\ZBp|^2$ \eqref{eq:cZB}, 
	where three diagrams with the photon coupling to the $u$-quarks are omitted as they are completely analogous to the 
	$b$-quark  ones.}
	\label{fig:dia-denom}
\end{figure}
\end{centering}

The LSZ-factor \eqref{eq:JBp} can be extracted from the following diagonal correlation function 
 \begin{eqnarray}
 \label{eq:C}
C(p_B^2, \momB^2) &=& i  \int_x e^{i x \rr}  \matel{ \phiB( \momB) }{ T \Hpz(x)  \Hpzda(0)  }{ \phiB( \momB)}   \nonumber \\
&=&  \int \frac{ds}{2\pi i} \frac{  \text{disc}_s C(s, \mphiB^2   ) }{s-p_B^2-i0} =   \frac{ |\ZBp|^2    }{ m_B^2- 
p_B^2} + \dots \;,
 \end{eqnarray}
where, as before, the dots stand for higher states and this time we do not show the LSZ procedure 
for the $\phiB$-particle explicitly, as it is straightforward.  The  quantity of interest is then determined  from
\eqref{eq:C}
\begin{equation}
\label{eq:cZB}
|\ZBp|^2  =   \lim_{p_B^2 \to m_B^2} (  m_B^2- p_B^2) C(p_B^2, m_B^2) \;.
\end{equation}
The computation proceeds in the same way as for the main process and 
the diagrams are shown in \FIG\ref{fig:dia-denom}.
One may be concerned as to 
whether the correlation function (or its discontinuity) are IR-finite as the $\phiB$-particle is on-shell. 
Fortunately, $|\ZBp|^2$ 
turns out to be IR-finite  and this follows from a physical argument. 
We may interpret $|\ZBp|^2$, via the optical theorem, 
 as an inclusive decay rate of  $\phiB(p_B) \to \bar b q  X_0(r)$ induced by the 
hypothetical effective Lagrangian $``{\cal L}_{\text{eff}} = \Hpz X_0 "$ where $X_0$  
is a neutral particle (of momenta  $r$,  which decouples when $r \to 0$).  Hence, by virtue of the KLN-theorem, 
which is based on  unitarity, we know that its discontinuity must be IR-finite (also in the  $m_u \to 0$ limit). We have checked that this is true by an explicit computation.  Once more, it is important that one takes the sum of all cuts as individual cuts are IR divergent.  

\subsection{Summary of the basic interpolating operator approach}
\label{sec:summary}

In summary, since the numerator reproduces all  universal IR-sensitive logs 
and the denominator is IR-finite, this strongly suggests that the proposed procedure is correct. 
In particular, the IR-finiteness of $\ZBp$ means that the expression \eqref{eq:master} has ``forgotten"
about its interpolating operator, as required, since it is an auxiliary in the LSZ formalism.  
The incorporation for several particles is straightforward  from the viewpoint of the interpolating operators; one can add as many as one desires to.  The same applies to non-scalar particles; 
for a proton  one  adds a scalar $\Phi_P$ and not a spin $\frac{1}{2}$-particle. 
The $\ORD(E_\ga^0)$ term in Low's theorem \eqref{eq:Lowthm}, related to spin,  
 is reproduced from the spinor 
in the form factor decomposition and is entirely kinematical.

\section{Relation to Non-local Operators} 
\label{sec:Dirac}

There are gauge  invariant formulations of QED, which are functionals of the photon field.
 These date back to the work of Dirac in 1955 
\cite{Dirac:1955uv}\footnote{This formulation has been taken up by lattice groups \cite{Lucini:2015hfa} 
with $C^*$-boundary conditions, originally introduced for studying monopole condensation
\cite{Polley:1990tf}, since the usual  periodic  boundary conditions are not compatible with it.}
and Mandelstam in 1962 \cite{Mandelstam:1962mi} where the photon field is  
integrated over spacetime and a spacelike path respectively. 
We first review the Dirac dressing in  \SEC\ref{sec:synopsis}, including the concept of dual gauges 
in \SEC\ref{sec:inter}.  In  \SEC\ref{sec:DiracphiB} we 
 show how to embed our approach at $\ORD(\al)$ into it.  In \SEC\ref{sec:iterated},  the all order equivalence is established, going beyond the Dirac dressing, using iterated integrals.

\subsection{Dirac Dressing}
\label{sec:synopsis}

One may introduce a gauge (compensating) factor  $U^{(P)}_\JDirac(x)$  
\begin{equation}
	\label{eq:DiracWL}
U^{(P)}_\JDirac(x)   \equiv e^{i Q_P \int d^4 y A_\mu(y) \JDirac^\mu(x-y)} \;,\quad \partial \cdot \JDirac = \de^{(4)}(x) \;,
\end{equation}
which is a functional of the photon field $A$ and a current $\JDirac$. 
The latter has  no direct relation 
to the electromagnetic  current but it is required to  satisfy the differential equation above 
and vanishing boundary condition   at infinity. 
There are many solutions to this equation and that will be the point of discussion soon. 
First, let us observe that 
\begin{equation}
{\psi}_\JDirac(x) \equiv U^{(\psi)}_\JDirac(x) \psi(x) \;,
\end{equation}
is gauge invariant since the gauge transformation \eqref{eq:GT}
\begin{equation}
\label{eq:GI-Dirac-operator}
U^{(\psi)}_\JDirac(x) \to e^{- i Q_\psi \la(x)} U_\JDirac(x) \;, \quad \psi(x) \to e^{i Q_\psi \la(x) } \psi(x) \;,
\end{equation}
 of the gauge factor and the fermion (or any other field) act to compensate each 
other.   In effect, the gauge flux is transported by $U^{(\psi)}_\JDirac$ to infinity where it 
is assumed not to matter (implicit by the imposed boundary condition).

\subsubsection{Dual gauges}
\label{sec:inter}

It is now immediate to define a gauge invariant version of $J_B$ \eqref{eq:JB}
\begin{equation}
	\label{eq:JBj}
\JBU(\JDirac,x)  \equiv  {\bar u_\JDirac} i \ga_5  b_\JDirac(x) = J_B(x) U^{(B)}_\JDirac \;,
\end{equation} 
by replacing the gauge variant quarks by gauge invariant ones. 
(Note that $U^{(B)}_\JDirac \to e^{i \la Q_B} U^{(B)}_\JDirac$). Hereafter, we suppress the 
subscript $(B)$ on the gauge factor for brevity.   
The operator $\JBU(\JDirac,x)$ is 
now a functional of $\JDirac$. This raises the question of whether specific choices are more convenient 
than others and or even more legitimate. Let us first set aside the latter point. 
It turns out that for a given $\JDirac$, one can often choose a gauge for which the gauge factor becomes 
trivial i.e. $U_\JDirac =1$. We may think of this in terms of the following equivalence or duality 
\begin{equation}
\label{eq:duality}
 \JBU(\JDirac_{\textrm{gauge}})  \to  J_B|_{\textrm{gauge}} \;,
\end{equation}
where ``gauge" is now an index for a specific current $\JDirac$ and the subscript after the vertical bar  
on the right hand side indicates
that the computation is to be performed in the specific gauge.  This calls for examples, 
for which we will choose the Coulomb  and Lorenz gauges (for the further example of the axial gauge, we refer the reader to  \cite{Lucini:2015hfa} where these aspects are nicely discussed).
\begin{itemize}
\item \emph{The Coulomb gauge:}
here, the current $\JDirac$, satisfying the differential equation \eqref{eq:DiracWL}, is 
 \begin{equation}
 \label{eq:Cgauge}
 \JDiracB{0}{\text{Coulomb}}(x) = 0 \;, \quad  \JDiracB{k}{\text{Coulomb}}(x) = - \de(x_0) \partial^k\pot(\vec{x}) \;, \quad \vec{\partial}^{\,2} \pot(\vec{x})= \de^{(3)}(x)\;,
 \end{equation}
($k = 1,2,3$)  where $\pot(x)$ is the solution of the differential equation compatible with 
the boundary condition for  $\JDirac$. Indices are interpreted as Minkowski ones and 
$\vec{x} \cdot \vec{y} = \sum_{i=1}^3 x^i y^i$.
 The gauge factor may be integrated by parts to 
 \begin{equation}
  U_{\JDiracO{\text{Coulomb}} }(x) = e^{ i Q_{B}\int d^3 y \,\vec{\partial}\cdot {\vec{A}}(y) \pot( \vec{x} - \vec{y} )} |_{ \vec{\partial}  \cdot {\vec{A}} = 0}  
    \to  1   \;,
 \end{equation}
 and  trivialises in the Coulomb gauge $ \vec{\partial} \cdot {\vec{A}} = 0$. 
 \item \emph{The Lorenz gauge:}
in this case, the current $\JDirac$, satisfying the differential equation, is 
 \begin{equation}
 \label{eq:CLorenz}
  \JDiracB{\mu}{\text{Lorenz}}(x) =  \partial^\mu \pot({x}) \;, \quad  
  \partial^2 \pot({x})= \de^{(4)}(x)\;.
 \end{equation}
Integrating by parts,  the gauge factor reads  
 \begin{equation}
 U_{\JDiracO{\text{Lorenz}}} (x)  =  e^{  i Q_{B}\int d^4 y \,{\partial}\cdot {{A}}(y) \pot(x-y)} |_{ \partial \cdot A= 0}
    \to  1 \;,  
 \end{equation}
and   trivialises in the Lorenz/Landau gauge, $ \partial \cdot A = 0$.\footnote{Since the gauge fixing term reads ${\cal L}_{ \mathrm{gauge} } =\frac{1}{2 \xi} (\partial \cdot A)^2 $, the Landau gauge $\xi = 0$ and the Lorenz gauge condition are equivalent (at least in perturbation theory).}
 \end{itemize}
 We wish to stress that the choice of gauge here is nothing but a computational trick or a matter of convenience.  The element of complexity in the gauge factor $U_{\JDiracO{\text{gauge}}}$ is moved 
 into computing with the gauge variant  operator $J_B$ in a specific gauge.  These two effects 
 of gauge dependence act to compensate each other.
 
 Let us now return to the question, alluded to  before, of whether  all choices
 of $\JDirac$ are equally valid. 
 We would think that the answer to this question ought to be yes in approaches 
 with an exact LSZ formula.  
 However, if the LSZ formula is approached in the sense of duality as in QCD sum rules 
 (cf. \APP\ref{app:QCDSR}), this is 
 not the case as it turns out that neither the Coulomb nor the Lorenz gauge current reproduce  
 the universal IR-logs in \eqref{eq:leptonic-rate}. This was  explicitly  verified using   the 
 gauge factor $U_{\JDirac}$.

\subsubsection{The $\phiB$-particle in Dirac dressing}
\label{sec:DiracphiB}

Hence, the natural question is whether our approach which is gauge invariant can be captured in this formalism with a specific current $ \JDirac $. The following expression achieves this task
 \begin{equation}
 \label{eq:Ourgauge}
   \JDiracB{\mu}{\Phi_B}(x) =  (\partial- 2 ip)^\mu e^{ixp} \pot(x)  \;, \quad ( \partial^2 +m_B^2) \pot(x) = \de^{(4)}(x)\;,
 \end{equation}
 where $p \equiv p_\phiB$ for brevity and  on-shell momentum   ($p^2 = m_B^2$).  
 As a solution to the differential equation \eqref{eq:DiracWL} with the appropriate boundary condition, 
 the Feynman propagator $ \pot(x) =  i \Delta_F(x,m_B^2)$ is chosen.
The gauge factor integrates by part to \begin{alignat}{2}
 \label{eq:Uour}
&  U_{\JDiracO{\Phi_B}}(x,p) &\;=\;&  e^{  i Q_{B}\int d^4 y \, e^{i(x-y)p} 
\Delta_F(x-y,m^2_B) (i \partial + 2p) \cdot {{A}}(y) }  \;,
 \end{alignat}
 a most familiar form. Namely, the exponent becomes the Feynman rule for scalar QED with the scalar being 
our $\phiB$-particle! 
In fact, a hint of this possibility was given by the Lorenz gauge case \eqref{eq:CLorenz} which, however, corresponds to the massless propagator with zero momentum insertion ($p_\mu =0$). Here, we have in effect extended this mechanism to the massive propagator with a non-zero momentum. 
For clarity, let us quote the corresponding interpolating operator 
\begin{equation}
 \label{eq:HpU}
 \JBU(x,p) \equiv J_B(x) U_{\JDiracO{\Phi_B}}(x,p) \;,
 \end{equation}
 where the superscript (D) stands for Dirac. 
 A natural question, in   view the discussion in  \SEC\ref{sec:inter}, is whether 
 there exists a dual gauge (that trivialises the gauge factor \eqref{eq:Uour})?
The answer is yes, 
\begin{equation}
\label{eq:Ourgauge-trivial}
 U_{\JDiracO{\Phi_B}}(x,p)|_{(i \partial+2p) \cdot {{A}} = 0}  \to 1 \;,
 \end{equation}
which is a peculiar axial gauge for which the photon propagator in momentum space assumes the form 
 \begin{equation}
 \label{eq:Delaxial}
  \Delta_{\mu\nu}(k)\Big|_{\Phi_B -\textrm{gauge}} = \frac{1}{k^2}\left(- g_{\mu \nu} - 
 n^2   \frac{k_\mu k_\nu}{(n \cdot k)^2} + \frac{k_{ \{ \mu} n_{\nu\}} }{ n \cdot k}  \right) \;, \quad n =  k + 2p \;,
 \end{equation}
with $k_{ \{ \mu} n_{\nu\}} = k_\mu n_\nu +  k_\nu n_\mu$, and $n^\mu  \Delta_{\mu\nu} =0$ as required.

\subsection{Iterated integral approach} 
\label{sec:iterated} 

It is clear that the form in \eqref{eq:Uour} is not suitable for higher order computations, or already the $\phiB$ self-energy 
correction.  
Matters can be improved by writing an expression with \emph{iterated} integrals.
For that purpose, let us define the following kernel 
\begin{equation}
\label{eq:KphiB}
\KB(z,y) \equiv  i Q_{B}  \, e^{izp}  \Delta_F(z,m^2_B) (i \partial+2p) \cdot {{A}}(y)  \;,
\end{equation}
suppressing the $p$ and $A$ dependence in $\KB$.  Then, the  improved and final version  reads 
 \begin{equation}
 \label{eq:HpV}
 \JBV(x,p) \equiv J_B(x) V_{\JDiracO{\Phi_B}}(x,p) \;,
 \end{equation}
 where
 \begin{equation}
 \label{eq:Vour}
  V_{\JDiracO{\Phi_B}}(x,p) = 1 +  \sum_{n \geq 1}   \int d^4 y_1 \dots d^4 y_n 
\KB(x-y_1,y_1) \dots \KB(y_{n-1}-y_n,y_n)\;,
 \end{equation}
 consists of the iterated kernels $K$.  In essence, this formula is the Dyson series for the $\phiB$-particle 
 where the propagators have been contracted already and this is the reason why the  $1/n!$-factor 
 from the exponential  has disappeared! To establish  the gauge transformation, 
$ V_{\JDiracO{\Phi_B}}(x) \to e^{i Q_B \la(x)} V_{\JDiracO{\Phi_B}}(x) $, by direct computation is not a simple matter. 
However, it is clear that it must hold since the $\phiB$-formalism is gauge invariant. 
In addition,  we have verified
the transformation law explicitly up to fourth order, witnessing intricate cancellations.

  The $ \JBV$  interpolating operator  is the most transparent and most generally valid version obtained in this paper as it clarifies a number of questions.
 For example, does the $\phiB$-particle need to be included into the running of the fine structure constant  $\al$? 
 The answer is negative since it   just ``lives" inside the factor $V_\JDirac$ and 
 does not figure in  the Lagrangian of the theory. This means that there is no coupling to charged 
 fermions other than through the interpolating operator itself.   As such it serves as a justification for the 
 rules applied previously. In summary we thus have the following relation
 \begin{equation}
 {{\cal Z}_B}= \ZBp  =  {{\cal Z}_B^{(D)}} + \ORD(\al^2) \;,
 \end{equation}
 formulated in terms of the respective LSZ-factors.\footnote{The $ \JBV$-formulation 
is  related to the  coherent state framework in the sense that the 
soft logs (not the hard-collinear log) are reproduced.
This is achieved by  taking the coherent state 
function to be the eikonal factor $\omega_\mu \propto p_\mu/(p \cdot k)$ which defines the coherent state  
$\state{\omega} \propto \exp(  \int d \Phi_\ga \omega^\mu a^\dagger_\mu)\state{0} $ with $a^\dagger_\mu$ 
being the photon creation operator 
\cite{Kulish:1970ut} (and  \cite{Zwicky:2021olr} for a more complete set of references).
Again, this has to be the case since soft resummation is equivalent to the coherent state approach at the leading log level.}

\section{Conclusions and Discussions}
\label{sec:conclusions}

In this work,  a method was proposed for incorporating  charged  hadrons 
via  gauge invariant interpolating  operators for perturbative methods such as QCD sum rules.
Technically, this  consists of adding the long distance field
$\phiB$ to the gauge variant operator $J_B$ \eqref{eq:JB}, that 
is  $\Hpz  \equiv J_B  \phiB$ \eqref{eq:JBp}. 
Intuitively, $\phiB$ takes on the r\^ole of the long distance $B$-meson splitting into its two valence partons $\bar u$ and $b$, thereby resolving the dilemma
 that bound states (e.g. hadrons) are beyond perturbation theory,
but essential for infrared-sensitive physics.  
Formally, $\phiB$ solves  two linked problems at once: $J_B \to \Hpz$ becomes gauge invariant since it is charge neutral, and 
the universal logs (cf. \eqref{eq:leptonic-rate} and the end of \SEC\ref{sec:main}) are reproduced.\footnote{For inclusive enough quantities, which are free of IR-logs, one may not need to use gauge invariant interpolating operators since the on-shell cuts will render it gauge invariant.   
This happens for $\Delta m = m_{B^+} - m_{B^0}$ type quantities \cite{Rowe:2023jlt}.}  
The main formula for computing processes is given in \EQ\eqref{eq:master}, where 
both parts, the numerator and denominator, are computed separately.  The modified LSZ-factor  $\ZBp$ is gauge invariant and  IR-finite which can be argued to hold on grounds of the  KLN-theorem. Together with the reproduction of the universal IR-logs, this consists of the cornerstone in validating the approach.

In \SEC\ref{sec:Dirac}, we established the link of the method to the Dirac dressing of charged fields
to   $\ORD(e^2)$, $\JBU$,   which can be found in 
\EQs \eqref{eq:HpU}  and \eqref{eq:Uour}. An improved version $\JBV$, valid to all orders,    
generalising the Dirac dressing by using  iterated  integrals has been given in \EQs\eqref{eq:HpV}
and \eqref{eq:Vour}.  
Reassuringly, this generalisation makes it clear that the $\phiB$-particle 
does not contribute to the running of 
the fine structure  constant $ \al $ since it does not appear in the Lagrangian of the theory.
The dual gauge, trivialising the gauge factor, has been identified as a peculiar axial gauge,  cf. 
\EQs \eqref{eq:Ourgauge-trivial} and \eqref{eq:Delaxial}.

Note that  the method generalises to any number of particles and any types of spins with remarks 
at the end of  \SEC\ref{sec:summary}.
One can add for each charged particle an operator of the form \eqref{eq:HpV}. However, 
one can use the trick of the trivialising gauge only once. 
Explicit results of the computation for leptonic decays, which necessitate one interpolating operator only, 
including numerics, are to follow in a forthcoming publication \cite{NRZ22}.

\acknowledgments

We would to thank Martin Beneke, 
Luigi Del Debbio, Matteo Di Carlo, 
Giulio Falcioni, Einan Gardi,  Max Hansen, Anton Ilderton Agostino Patella, Antonin Portelli 
and further participants of the ``QED in Weak Decays" workshop in Edinburgh 
for useful discussions and feedback. In particular, we acknowledge the input of Matt Rowe to this work
in terms of proofreading, discussions and computations
RZ is supported by an STFC Consolidated Grant, ST/P0000630/1. SN is supported by the GLUODYNAMICS project funded by the ``P2IO LabEx (ANR-10-LABX-0038)" in the framework ``Investissements d’Avenir" (ANR-11-IDEX-0003-01) managed by the Agence Nationale de la Recherche (ANR), France.

\appendix

\section{On the Necessity of Gauge Invariant Interpolating Operators}
\label{app:necessity}

In this appendix, we comment under which circumstances interpolating 
operators for charged mesons are necessary or not.  This seems mandatory since
processes with charged hadrons have been considered in the literature using gauge variant  
interpolating operators. This includes i) $\lepB \ga$ where $\ga$ is a hard photon, to be discussed in \APP\ref{app:hard}, and
 ii) leptonic decays in  lattice Monte Carlo simulations, to be discussed in \APP\ref{app:exact}.

\subsection{Relation to computations of $\lepB \ga$ with a hard photon}
\label{app:hard}

There is some good tradition in using QED gauge dependent 
interpolating operators $J_B$ \eqref{eq:JB} for  $\lepB \ga$  in QCD sum rule approaches; 
e.g at $\ORD(\al_s^0)$
\cite{Ali:1995uy,Khodjamirian:1995uc} 
and $\ORD(\al_s)$ \cite{Janowski:2021yvz} or for the 
$g_{B^-B^{*-}\ga}$ coupling at $\ORD(\al_s^0)$ 
\cite{Khodjamirian:1995uc} and  $\ORD(\al_s)$ \cite{Pullin:2021ebn}.
This raises the obvious question of how the  issues raised, at the beginning of the paper, were avoided.

\begin{centering}
\begin{figure}[h!]
\begin{center}
\includegraphics[width=1.0\linewidth]{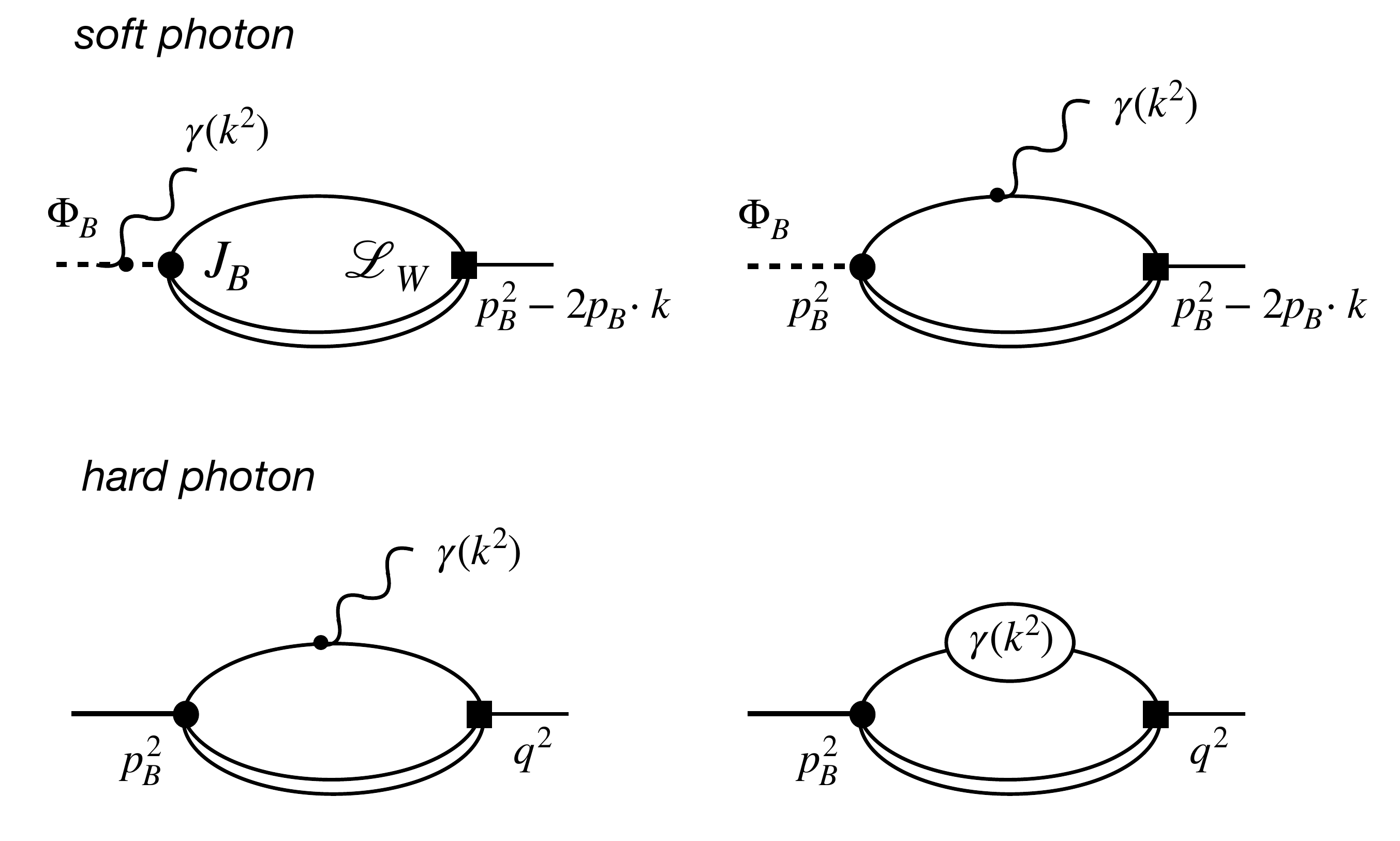}  
	\caption{\small  Comparison of soft and hard photon approaches. In the soft photon approach 
	(top diagrams), $q^2 \to p_B^2 - 2 p_B \cdot k$ (since $k^ 2=0$) 
	and  $p_B \cdot k$ becomes an independent 
	variable.  
	Top left (right) diagram correspond to the $\phiB$ (quarks)-term in \eqref{eq:PiPP}. 
	Note that  the top left diagram reproduces the Low term and 
	the top right diagram is finite as $p_B \cdot k \to 0$. 
	In the hard photon approach (bottom diagrams), $q^2$ is an independent variable. The Low term
	is produced by the bottom left diagram (cutting in $p_B^2$), 
	while the bottom right diagram probes the structure. 
	In particular, the blob is the photon distribution amplitude (DA) and is 
	generated by a sum over intermediate mesons $\rho^0,\omega,\phi$ 
	with quantum numbers of the photon $J^{PC} = 1^{--}$.
	For an elaborate discussion of this viewpoint, see \APP A of \cite{Albrecht:2019zul}. 
	The mappings of the diagrams between the two approaches are as follows: top left $\subset$ 	
	bottom left, bottom right 	$\subset$ top right. The first relation follows from the fact that 
	the point-like interaction on the top-left does not ``know" anything about the non-perturbative $\rho-\ga$ conversion that the photon DA represents and this also implies the second relation. }
	\label{fig:triangle}
	\end{center}
\end{figure}
\end{centering}

First and foremost,  at $\ORD(e)$, that is, for a single photon emission, these observables  are 
formally analogous  to $\lepB \rho^0(k)$ and $g_{B^-B^{*-}\rho^0}$  respectively. 
The main point is that the $\rho^0$ 
or the hard photon (say  $E_\ga \geq \Lambda_{\text{QCD}}$) are considered as  separate particles
and formally, this implies that  $\{ q^2, p_B^2,k^2\} $ 
are the independent kinematic variables, referred to as the hard photon approach.\footnote{  
In the soft photon approach, pursued in this work, $\{ p_B \cdot k, p_B^2,k^2\} $ are the independent variables. More comments are to follow further below.}
The situation is  illustrated in  \FIG\ref{fig:triangle} with further comments in the caption. 

Computing $\ORD(\al_s)$ corrections to 
$\lepB \ga$ and $g_{B^-B^{*-}\ga}$  is demanding, but straightforward,  because $J_B$ 
is QCD gauge invariant. 
However, complications arise  if we were to compute $\ORD(\al)$ radiative corrections  
as  $J_B$ is QED gauge variant. 
In that case, the introduction of the gauge invariant interpolating operator becomes, in our opinion, a necessity
in perturbation theory.  One should regard $\lepB \ga$ with the hard photon as a LO process, and it is only its radiative corrections that necessitate the introduction 
of the soft photon (to complement the virtual corrections). This is in line with the picture of coherent states
(e.g.  \cite{Zwicky:2021olr} and relevant references therein).

It is still an interesting question as to how $\lepB \ga$ at $\ORD(e)$ with gauge invariant interpolating 
operator ${\cal J}_B$ reduces to the case of the gauge variant operator $J_B$; in particular, of how
 the Low terms  emerge,  giving rise to the universal IR-logs for which we  had to introduce 
 the $\phiB$-particle.  
\emph{Low's theorem} \cite{Low:1954kd}, quoted with the same convention as in  \cite{Zwicky:2021olr},
 states that  adding a real photon to a transition $\al \to \be$
\begin{equation}
\label{eq:Lowthm}
\matel{\be \ga(k,\la) }{ S} {\al} 
= \frac{c_{-1}}{E_\ga} + c_0  + c_1 E_\ga + \ORD(E_\ga^2) \;,
\end{equation}
the two first 
terms in an $E_\ga$-expansion are universal and given by
\begin{equation}
\frac{c_{-1}}{E_\ga}  =  \sum_{j} \hat{Q}_j \frac{ \eps^*(k,\la) \cdot \hat{p}_j    }{ k \cdot \hat{p}_j - i0}
\matel{\be }{ S} {\al} 
 \;, \quad 
c_0 = -i \sum_{j} \hat{Q}_j \frac{ \eps_\mu^*(k,\la)  k_\nu J_j^{\mu \nu}    }{ k \cdot \hat{p}_j - i0} 
\matel{\be }{ S} {\al} \;.
\end{equation}
Above, $J_j^{\mu\nu}  = i \hat{p}_j^{[\mu} \partial^{ \nu]}_{\hat{p}_j}$ is the orbital angular momentum operator (and square brackets denote anti-symmetrisation). Hatted quantities are plus(minus) its value for outgoing(incoming) particles.

 To further simplify matters, we consider the S-P interaction 
\eqref{eq:LW},
in which case there are no $B \to \ga$ form factors by helicity conservation 
 and it all reduces to the Low-term corresponding to the emission from the $B$-meson 
(and the charged lepton of course).  
More concretely, at LO, the amplitude  factorises 
\begin{alignat}{2}
&  {\cal A}^{S-P}_{\lepB \ga} &\;=\;&
\matel{\ga \ell^- \bar \nu }{- {\cal L}_{ \lepB}^{\small (S-P)}}{B^-} \nonumber  \\[0.1cm]
& &\;=\;&-g_{S-P}\matel{\ga }{ \bar u (1\mi \ga_5   )b}{B^-}  \matel{\ell^- \bar \nu }{ \ell (1\mi \ga_5   ) \nu}{0}  + 
\dots\;,
\end{alignat}
where  $g_{S-P}$ is an irrelevant constant for our purposes and 
the dots stand for photon emission from the lepton.  
We focus on the first term which, by  Low's theorem,   gives
\begin{equation}
\label{eq:Low}
 {{\cal A}}^{S-P}_{B \to \ga} \equiv  - i m_b \matel{\ga }{ \bar u (1\mi \ga_5   )b}{B^-(p_B)}  
= -   e  Q_B \frac{p_B \cdot \eps^*}{p_B \cdot k} {\cal A}_{LO}  
= - \frac{ 2 e Q_B  m_B^2 f_B  }{m_B^2 -q^2} (p_B \Cdot \eps^*) \;,
\end{equation}
where  ${\cal A}_{LO} =  m_B^2 f_B$  (having set $m_u = 0$), $p_B \cdot k = \frac{1}{2}(m_B^2 -q^2)$, 
$\eps$ is the photon's polarisation vector, 
 and by parity, only the $\ga_5$-part contributes.  Two remarks are in order: i) as this is an on-shell matrix element, hence  
 $p_B^2 = m_B^2$ ii)  the expression \eqref{eq:Low} is exact, as stated  before. 

Now, we 
would like to see how this works out in
our approach (with gauge invariant interpolating operator). The correlation function 
\eqref{eq:mainSR} evaluates to 
\begin{equation}
\label{eq:PiPP}
\Pi^{(\ga)}(p_B^2, \momB^2,q^2)  \propto e Q_B
 \left( \left[ \frac{\Pi_P(q^2)}{m_B^2 - q^2}   \right]_{\phiB} + 
\left[ \frac{\Pi_P(p_B^2)- \Pi_P(q^2)}{p_B^2 - q^2}  \right]_{ \mathrm{quarks} }   \right) (p_B \Cdot \eps^*) + \dots \;,
\end{equation}
where  the dots once more stand for the photon emission from the lepton 
(and $(p_B - \momB) \cdot \eps^*$-terms which have to be dropped as they are unphysical). 
The first term in the $p_B \Cdot \eps^*$-structure corresponds to emission from the $\phiB$-particle, the second term to  emission  from the quarks  
and $\Pi_P(s)$ is  the following 2-point  function
\begin{equation}
\label{eq:PiP}
\Pi_P(p^2) =   i  \, m_b ^2 \int_x e^{ipx} \matel{0}{T J_B(x) J^\dagger_B(0)}{0} \;.
\end{equation}
which is further scrutinised  in \SEC\ref{app:QCDSR}.  Note that the  extra factor of $ J_B $ arises from the quark part of the weak Lagrangian.
The quark emission term in \eqref{eq:PiPP} can be established without computation as being the unique solution to the QED Ward identity. This  goes hand in hand with the statement that \eqref{eq:Low} is exact.
Now, we have all the information in order to investigate the differences between the two approaches. 
\begin{itemize}
\item \emph{Hard photon approach $\{ q^2,p_B^2,k^2 \}$}: in this case $q^2$ is considered an independent variable 
and the discontinuity in $p_B^2$ is given by
\begin{equation}
\label{eq:hard}
\text{disc}_{p_B^2} \Pi^{(\ga)}(p_B^2, \momB^2,q^2) \propto  \frac{\text{disc}_{p_B^2} \Pi_P(p_B^2)}{p_B^2-q^2}  \;,
\end{equation}
which reproduces the Low-term   \eqref{eq:Low} to the extent that $p_B^2 \to m_B^2$, which holds 
in QCD and approximately  for QCD sum rules at the level of quark hadron duality as noticed in \cite{Janowski:2021yvz} (and cf. \APP\ref{app:QCDSR}). The latter  is a sufficiently good approximation  as long as $q^2$ is not too close to $m_B^2$ which is the premise of the hard photon QCD sum rule  approach. 
\item \emph{Soft photon approach $\{ p_B\Cdot  k,p_B^2,k^2 \}$}: it is characterised  
by $E_\ga \ll \Lambda_{\text{QCD}}$ 
and it makes more sense to expand around $E_\ga$ (or better its covariant form $p_B\cdot  k$). 
\EQ\eqref{eq:PiPP}, omitting its arguments, then becomes 
\begin{equation}
\label{eq:PiPP2}
\Pi^{(\ga)} \propto  e Q_B
 \left( \left[ \frac{\Pi_P(q^2)}{2 \momB \Cdot k}   \right] + 
\left[ \frac{\Pi_P(p_B^2)- \Pi_P(q^2)}{2 p_B\cdot  k}  \right]   \right)\Bigg|_{q^2 \to p_B^2-2 p_B\cdot  k} 
 (p_B \cdot \eps^*) + \dots \;,
\end{equation}
and formally we traded $q^2$ for  $p_B\cdot  k$. 
Note that, at the end, $p_B\cdot  k$ is set equal to $\momB \cdot k =  E_\ga m_B$ 
 The first thing to notice is that in the $k \to 0$ limit, the second term becomes 
$\Pi'_P(p_B^2)$, the derivative of the $2$-point function, and contributes to $c_0$, 
but does not reproduce the $c_{-1}$-term in  \eqref{eq:Lowthm}.\footnote{This is in accordance with the 
Kinoshita-Poggio-Quinn-theorem (cf. \cite{Zwicky:2021olr}  for references) which states 
that in renormalisable theories, off-shell correlation functions are free from IR singularities for
non-exceptional momenta.} This r\^ole is reserved the $\phiB$-term! 
This highlights the necessity of introducing the $\phiB$-particle in this approach. 
Most importantly, 
the two terms in $\Pi_P(q^2)$ exactly cancel each other 
\begin{equation}
\label{eq:soft-approach}
\text{disc}_{p_B^2} \Pi^{(\ga)}|_{q^2 \to p_B^2-2 p_B\cdot  k} 
 \propto  \frac{\text{disc}_{p_B^2} \Pi_P(p_B^2)}{2 p_B \Cdot k }  \;,
\end{equation}
and the Low term appears in its exact form \eqref{eq:Low} (after the LSZ formula in $p_B^2$ is applied). 
\end{itemize}
The example of the S-P interaction almost appears a bit too simple to illustrate the point but 
in fact it is not. In the V-A case, there would simply be many other terms contributing to the structures in \eqref{eq:Lowthm} other than the Low-term (e.g. \cite{Janowski:2021yvz}).  

An interesting question that one could raise is the following:
how can taking cuts in $p_B^2$ in the hard photon approach be equivalent  to taking cuts in $p_B^2$ itself and 
an extra cut in $q^2 = p_B^2 - 2 p_B \cdot k$ in the soft photon approach? 
The answer is that for $s_0 - m_b^2 < m_B^2 - q^2$,  where $ s_0 $ 
is the (effective) continuum threshold, this extra cut does not actually contribute 
to the sum rule (cf.  \eqref{eq:ALO2}).  This translates to $q^2 < 14 \GeV^2$ for 
typical values of $s_0 = 35 \GeV^2$ and 
$m_b = 4.6\GeV$ (pole or kinetic scheme $m_b$-mass). 
This is what is usually assumed in the light-cone expansion indeed (e.g.  \cite{Janowski:2021yvz}). 
Hence for  $ q^2 > 14 \GeV^2$, or well-above, the hard photon approach 
gradually breaks   down and the soft photon variables become appropriate. 

\subsection{Exact  LSZ formula and gauge variant interpolating operators}
\label{app:exact}

Let us comment on the necessity of gauge invariant interpolating operators. 
If one aims only at $\ORD(\al)$ in a decay process   and 
one has an exact LSZ formula, then it would seem possible
to work with gauge variant interpolating operators for hadrons. 
This is because the hadrons are  the QCD hadrons (not corrected by QED) 
and those states are well-isolated in the spectrum.  
The Euclidean  correlation function then assumes the form 
\begin{equation}
 Z_B(gauge)  \times \text{amplitude} \times e^{- E_B t_E}  +  \dots\;,
\end{equation}
where the dots stand for exponentially suppressed 
terms (cf.~\eqref{eq:euclid}) and $Z_B$ corresponds to the gauge variant LSZ-factor of the $B$-interpolating operator in use. 
The exact LSZ formula emerges in the limit of infinite Euclidean time  separation of 
the meson source.  In this case, the exponentially suppressed terms disappear as $t_E \to \infty$ 
and the sole 
gauge dependence is in $Z_B$ and can be cancelled by  computing it from an appropriate correlation function.   This is the idea underlying 
the QED$_L$  \cite{Carrasco:2015xwa} and the QED$_{m_\ga} $\cite{Endres:2015gda} lattice approaches.

When aiming for corrections beyond $\ORD(\al)$, matters are more delicate since charged states are not isolated anymore (QED IR-problem, see  \cite{Zwicky:2021olr} for references) 
and an explicit gauge invariant formulation seems more appropriate. 
This is the idea
behind the $C^*$-boundary method \cite{Lucini:2015hfa}.  So far, this method has only been  
applied to hadronic masses and the specifics for (leptonic) decay rates have not been 
proposed to date.

\section{Brief review of the  LSZ formalism and QCD sum rule approach}
\label{app:QCDSR}

In this appendix, we briefly review the LSZ formalism (e.g. \cite{Duncan:2012aja}) itself 
and how it is handled in QCD sum rules \cite{Shifman:1978bx}.  We do so by considering the $2$-point function in \EQ\eqref{eq:PiP} relevant to $\lepB \ga$ for S-P interactions.
In QCD and perturbative QCD (pQCD), the most important terms for our considerations are
\begin{alignat}{2}
\label{eq:PiPex}
& \Pi_P(s)\Big|_{ \mathrm{QCD} }  &\; = \;&   \frac{m_B^4  f_B^2}{s-m_B^2} + \dots  \;, \\[0.1cm]
\label{eq:PiPex2}
& \Pi_P(s)\Big|_{ \mathrm{pQCD} }  &\; = \;& \frac{N_c}{8 \pi^2}  m_b ^2  \left(1- \frac{m_b^2}{s} \right)^2 s\, \ln( m_{b}^2 - s) 
+ \dots  \;,
\end{alignat}
where $N_c$ are the number of colours. The dots stand for higher states in the spectrum in 
\eqref{eq:PiPex} 
 and  non-logarithmic terms,  condensate terms and $\ORD(\al_s)$-contributions in \eqref{eq:PiPex2} 
 (for pQCD, 
 cf.   \cite{Pullin:2021ebn} for explicit results).  
In the  LSZ approach, repeating the steps in \eqref{eq:LSZ}, 
 one would extract the amplitude ${\cal A}_{ \mathrm{LO} }$, referred to below   \eqref{eq:Low}, by
\begin{equation}
\label{eq:ALO}
{\cal A}_{ \mathrm{LO} } =  \frac{1}{\ZB'} \lim_{p_B^2 \to m_B^2}(  m_B^2 \mi p_B^2)  \Pi_P(p_B^2)\Big|_{ \mathrm{QCD} } = m_B^2 f_B  \;.
\end{equation}
Here, $\ZB' = m_b \ZB =  m_B^2 f_B $  (cf. also  \eqref{eq:JB}) and the dots in \eqref{eq:PiPex} vanish as they do not have a pole in $(p_B^2 \mi  m_B^2)$.
 Let us consider this aspect by a dispersive representation (``s.t." stands for subtraction terms)
 \begin{equation}
  \Pi_P(p^2) = \int_0^\infty  ds  \frac{\rho_P(s)}{s- p^2 -i0}  + \text{ s.t.} \;,
  \end{equation}
 with $2 \pi i \rho_P(s)   =  \text{disc}_s \Pi_P(s)$. From \eqref{eq:PiPex}, one gets
 \begin{alignat}{2}
\label{eq:rhoPex}
& \rho_P(s)\Big|_{ \mathrm{QCD} }  &\; = \;&   m_B^4  f_B^2   \de(s-m_B^2) + \dots  \;, \nonumber \\[0.1cm]
& \rho_P(s)\Big|_{ \mathrm{pQCD} }  &\; = \;& \frac{N_c}{8 \pi^2}  m_b ^2  \left(1- \frac{m_b^2}{s} \right)^2 s\, \theta(s-m_B^2)   + \dots  \;.
\end{alignat}
Subtraction terms are eliminated by a Borel transform 
 which maps any polynomial in $s$ to zero and 
 $1/(s-p^2) \to \exp(-s/M^2)/M^2$ (with $M^2$  the Borel mass). 
We may then rewrite \eqref{eq:ALO}
\begin{equation}
\label{eq:ALO2}
{\cal A}'_{ \mathrm{LO} }[ \rho_P] =  \frac{1}{\ZB}  \int_{ \mathrm{cut} }^{s_0}  ds \, e^{(m_B^2-s)/M^2} \rho_P(s) \;,
\end{equation}
where ``cut" marks the start of the discontinuity and $s_0$ is just below the onset of the first excited states which are of the order
 of $(m_B + 2 m_\pi)^2$ to $(m_B +  m_\rho)^2$.  
 The difference between QCD and QCD sum rules is now most clear 
  \begin{alignat}{3}
\label{eq:ALOp}
& {\cal A}'_{ \mathrm{LO} } [ \rho_P\Big|_{ \mathrm{QCD} } ]   &\; = \;&  {\cal A}_{ \mathrm{LO} } \;,  \qquad & & \text{exact}   \;, \nonumber \\[0.1cm]
& {\cal A}'_{ \mathrm{LO} }  [ \rho_P\Big|_{ \mathrm{pQCD} } ]   &\; \approx \;&  {\cal A}_{ \mathrm{LO} } \;, \qquad & & \text{QCD sum rule} \;,
\end{alignat}
as it reduces to which density is in use. 
The approximation made can be quantified by\footnote{It is tempting to take the limit  
$M^2 \to 0$ as then the higher states would decouple. However, the problem with this is that then the operator products expansion does not converge in that case. $M^2 \to 0$ is in some sense the analogue of infinite Euclidean time separation in lattice QCD.}
\begin{equation}
\label{eq:semiglobal}
\int_{s_0}^{\infty}  ds  \,e^{(m_B^2-s)/M^2} \rho_P(s)\Big|_{ \mathrm{pQCD} } \approx 
\int_{s_0}^{\infty}  ds  \,e^{(m_B^2-s)/M^2} \rho_P(s)\Big|_{ \mathrm{QCD} }  \;,
\end{equation}
and is sometimes referred to as semi-global quark-hadron duality 
\cite{Shifman:2000jv}. In practice, it may be expected to hold to within $30\%$ and if the pole term dominates by $60\%$, this leads to an uncertainty of roughly $10\%$ \cite{Shifman:1978bx}. 
In practice, most sum rules 
are ratios of sum rules in fact, such as \eqref{eq:master}, and this effect cancels to a considerable extent.
\EQ\eqref{eq:semiglobal} can be expected to work well when the higher spectrum is broad, that is, if there are no further narrow resonances, which is most often the case. 

At last, it is worthwhile to sketch the analogue of the LSZ formula in Euclidean field theory in which
 the positive frequency correlation function in  the  time-momentum representation is considered
\begin{equation}
\label{eq:euclid}
\Pi^{+}_{P}(t_E,\vec{p}^{\,2}) =  
m_b^2 \int d^3 x\, e^{i \vec{x} \cdot \vec{p}}   \matel{0}{J_B(x) J^\dagger _B(0)} {0}
\propto  (\ZB')^2  \, e^{- E_B t_E}   +  \dots\;.
\end{equation}
Above the dots stand for exponentially suppressed terms and   $\ZB'$ has been defined below \eqref{eq:ALO}. 
The physical matrix element emerges in  the $t_E \to \infty $ such that the suppressed terms disappear.

\bibliographystyle{utphys}
\bibliography{../../Refs-dropbox/References_QED.bib}

\end{document}